\nonstopmode
\documentstyle[12pt,aasms4]{article}
\begin{document}
\title{Confirmation and Analysis of Circular Polarization from Sagittarius A*}
\author{R.J. Sault}
\affil{Australia Telescope National Facility, P.O. Box 76, Epping, NSW, 1710,
Australia}
\authoremail{rsault@atnf.csiro.au}
\and
\author{J.-P. Macquart}
\affil{Research Centre for Theoretical Astrophysics, 
School of Physics, University of Sydney, NSW, 2006, Australia}
\authoremail{jpm@physics.usyd.edu.au}

\begin{abstract}
Recently Bower et al. (1999b)\markcite{bower99b}
have reported the detection of circular
polarization from the Galactic Center black hole candidate, Sagittarius A*.
We provide an independent confirmation of this detection, and provide
some analysis on the possible mechanisms.
\end{abstract}


\section{Introduction}
The flat-spectrum, compact radio source in the Galactic Center, Sagittarius A* (Sgr~A*), has
long been believed to mark a massive black hole at the dynamical center
of the Galaxy (Lynden-Bell \& Rees 1971)\markcite{lynden71}.
Eckart \& Genzel (1996,1997)\markcite{genzel96}\markcite{genzel97}
and Ghez et al. (1998)\markcite{ghez98} provide compelling evidence
that there is a dark mass of $\approx 2.6\times 10^6\,M_\odot$ coincident
with Sgr~A*.
Furthermore VLBI observations suggest the intrinsic size of Sgr~A* is 
no larger than 1 to 3.6~AU (Krichbaum et al. 1998; Lo et al. 1998; Rogers 
et al. 1994; Bower \& Backer 
1998)\markcite{krichbaum98}\markcite{lo98}\markcite{rogers94}\markcite{bower98}.  
The observed size, however, is substantially
larger as a result of scattering by interstellar electrons in the vicinity
of Sgr A* (Davies, Walsh \& Booth 1976)\markcite{davies76}, and follows a $\lambda^2$
dependence.

If Sgr~A* is a synchrotron source, polarized emission may be be expected, and 
would prove a tight constraint on some of the proposed
models. However, despite numerous attempts, linear polarization has not been
detected from Sgr~A* (e.g. Bower et al., 1999a)\markcite{bower99a}.
This is so even at high frequencies,
or with experiments which should be sensitive to linear polarization with
high Faraday rotation measures (Bower et al. are sensitive to limits of
RM = $10^7$ rad m$^{-2}$).

Recently Bower et al. (1999b)\markcite{bower99b} have reported the detection of circular
polarization from Sgr~A* at 4.8 and 8.4 GHz. We report an independent
confirmation of this detection at 4.8 GHz using the Australia Telescope
Compact Array (ATCA). We note that we have used a different
telescope, calibrators and calibration procedures and software to Bower et al.

Care is needed in analysing radio astronomy data for circular polarization,
particularly given that the observed level of circular polarization
in synchrotron sources is always small. A minor error in the polarimetric calibration can
allow a fraction of total intensity to masquerade as circular polarization.
Such a miscalibration will lead to an erroneous image which
has no obvious error artifacts.
This is unlike observations of linear polarization
with alt-az antennas (small miscalibration will lead to
artifacts, rather than an apparently clean image).
The VLA's off-axis design and circularly-polarized feeds also make it
a poor instrument for circular polarization measurements. Given these
caveats, and the general mixed history of the detection of circular polarization,
we believe an independent confirmation adds significant weight
to the detection of Bower et al.

\section{Observations and Results}

We have made one new observation and used two archival observations
from the ATCA to independently test
the detection of Bower et al. The ATCA is a radio interferometer situated
in eastern Australia at a latitude of $-30^\circ$. It consists of
6 antennas over a 6~km baseline. With an on-axis
feed design, dual linear polarimetric measurements and alt-az
antenna mounts, it is an excellent instrument for the measurement of
circular polarization.  Three observations, made in March 1996, 1997
and 1999, were analysed.
The observations were 12~h, 8~h and 6~h in length, respectively.
These observations used a variety of bandwidths, array configurations and
correlator settings, but were all made at 4.8 GHz.
All runs included observations of the blazar PKS~B1730-130 approximately once
every 30~min and included at least one observation of the ATCA's primary
flux calibrator,  PKS~B1934-638.
With the ATCA having linearly polarized feeds, the most important calibration
step is in determining the antenna polarization leakage terms (the so-called
``D terms''). This was done using the PKS~B1934-638 data, which we have
assumed to have Stokes parameters of
$(I,Q,U,V) = (5.829,0,0,1.5\times10^{-3})$~Jy. PKS~B1934-638 is a 
GHz peaked spectrum source. Numerous ATCA and Parkes observations have
failed to detect linear polarization from it (even with appreciable rotation
measure). However work by Komesaroff el al. (1984)\markcite{komesaroff84}
and Rayner et al. (submitted to \mnras)
suggest some weak circular polarization. The value we adopt is from
Rayner et al. From the data for PKS~B1730-130, we have performed a
simultaneous solution for antenna gains as a function of time and
source polarization (note that PKS~B1730-130 is known to be time-variable
and circularly polarized at the level of several milliJanskys).
The reduced PKS~B1730-130 data was consistent with a time-variable
polarized point source.

In the polarimetric calibration process, we have included a subtle
geometric correction as follows.
Nominally, the geometry of the axes of all the ATCA antennas is
identical.
In reality, of course, this is not the case, and the deviation from
the nominal axis geometry, which is typically of order $1'$, is determined in
an antenna pointing solution.
For high precision work using alt-az antennas (such as the ATCA antennas) 
on sources that transit
near the zenith, the true antenna geometry needs to
be used in the calculation of parallactic angle (Sault et al, 1991\markcite{sault91};
Kesteven 1997\markcite{kesteven97}). With the declination of Sgr~A*
($\delta=-29^\circ$) differing from the ATCA's latitude by only $1^\circ$, 
this is a detectable effect.

The minimum spacing used in the analysis was 50~k$\lambda$ in total intensity
and 5~k$\lambda$ in circular polarization.  The total intensity limit is to
avoid confusion from the extended emission in the Galactic Center.  In circular
polarization, the only emission is from Sgr~A*, and so confusion is not an
issue.  However the use of a minimum baseline for circular polarization excludes
possible contamination from leakage of the rapidly rising total intensity
emission at short spacings.  This would be caused by small residual polarization
calibration errors.  Table 1 summarizes the results of our observations.
We give the total intensity, circular polarization and
fractional circular polarization of Sgr~A*. We also give the RMS
residual in the Stokes $V$ image and $\sigma_{\rm V}$ (the theoretical
noise in the Stokes-$V$ image which would result from the
measured receiver noise). The results show good self-consistency, and
agree well with the VLA detection of -2.0~mJy.

\placetable{Table1}

\section{Discussion}
The circular polarization properties of Sgr~A* are broadly consistent with 
those found in the cores of extragalactic radio sources.  The 
0.3 -- 0.4\% circular polarization of Sgr~A* is toward the high end  
of the range -- typical 
values for extragalactic objects are 0.05 to 0.5 \% 
(e.g. Roberts et al. 1975\markcite{roberts75}, de Pater 
\& Weiler 1982\markcite{dePater82} and 
Weiler \& de Pater 1983\markcite{weiler83}).  The absence of linear 
polarization is, however, unusual.

Variations in the circularly polarized flux 
indicate either a change in the intrinsic degree of circular 
polarization or that the circularly polarized source is small enough 
to exhibit the effects of interstellar scintillation.
Since the circular polarization in extragalactic sources is sometimes found to be 
variable (Komesaroff et al. 1984\markcite{komesaroff84}) and the total 
intensity of Sgr~A* is itself variable (e.g. Brown \& Lo 
1982\markcite{brown82}), it is of interest to 
place even a crude constraint on the degree of variability of the circular 
polarization.  Although obviously hampered by the small number of 
measurements, we note the 
possibility that the circularly polarized component is 
variable.  The normalized variance of the total intensity, defined by 
$\langle [I - \bar{I}]^2 \rangle / \bar{I}^2$, is 0.11.  The corresponding
quantity for the circular polarization is 0.16, however at least 0.09 
(60\%) of this may be attributed to measurement uncertainty.

The degree of circular polarization ($V/I$) may also vary.  Variation in this 
quantity implies that either the intrinsic circular polarization is 
variable or, if the source scintillates, 
that the polarized emission experiences different phase fluctuations 
along its ray path compared to the bulk of the (unpolarized) emission.
We can place a constraint on the variability of the degree of 
circular polarization: the $3\sigma$ upper limit  
$\Delta(V/I)/[\bar{V}/\bar{I}]$, is $\approx 25$\%.  This number is 
only relevant to the variations on the timescale comparable to our observing 
intervals (i.e. one year).

\section{Origin of the Circular Polarization}
It is of considerable interest to consider the physical properties of 
Sgr~A* which give rise to the observed circular polarization.  It 
is possible that the circular polarization is intrinsic to the synchrotron
emission (Legg \& Westfold 1968\markcite{legg68}), or it may result 
from one of several propagation-related mechanisms: `circular 
repolarization' converts linear to circular polarization and may occur 
either in a cold plasma 
(Pacholczyk 1973\markcite{pacholczyk73}), or in an electron-positron 
pair dominated 
plasma (Sazonov 1969\markcite{sazonov69}, Jones \& O'Dell 
1977a,b\markcite{jones77a,jones77b}).  Circular polarization may 
also be induced by 
scintillation (Macquart \& Melrose 1999\markcite{macquart99}).

It is possible that the circular polarization is
associated with only a small component of the total flux density of Sgr~A*.
In the following discussion we therefore denote the degree of circular 
polarization as $m_c = 0.0035 \, \xi$, where $\xi \geq 1$.  

The circular polarization due 
to sychrotron radiation from a power law distribution of relativistic 
electrons $N(\epsilon) \propto \epsilon^{-2 \alpha -1}$ is 
(Melrose 1971\markcite{Melrose71})
\begin{eqnarray}
m_c = \frac{\cot \theta}{3} \left( \frac{\nu}{3 \, \nu_H \sin \theta}
\right)^{-1/2} f(\alpha),
\end{eqnarray}
where $\theta$ is the angle between the line of sight, $\nu$ is in hertz 
and $\nu_H = 2.8 \times 10^6 B$~Hz is the electron gyrofrequency, where 
$B$ is the magnetic field in gauss.   The function $f(\alpha)$ is a weak 
function of the spectral index, $\alpha$; for 
optically thick emission in the limit of strong Faraday rotation 
$f(\alpha)$ only varies monotonically between 
0.6 and 2.0 for $\alpha$ between 0 and 2 (see Melrose 1971).
The observed flux density of Sgr~A* increases with frequency up to at 
least 850~GHz (\markcite{falcke98}Falcke et al. 1998, Serabyn et al. 
1997\markcite{serabyn97}) and is 
roughly proportional to $\nu^{1/3}$, suggesting that the source is 
optically thick at $\nu = 4.8$~GHz.  The high magnetic fields and 
particle densities thought to occur to occur in the source (e.g. Beckert 
et al. 1996\markcite{beckert96}) motivates 
the use of the strong Faraday rotation limit.  The high RM measurements 
(e.g., Yusef-Zadeh, Wardle \& Parastaran 1997\markcite{yusef97}) in the 
vicinity of Sgr~A* support the use of the strong Faraday rotation 
limit.  (The strong Faraday rotation limit does not necessarily imply linear depolarization and is applicable whenever 
negligible absorption occurs over a path length in which the plane of 
linear polarization rotates through 2~$\pi$ radians.)
The electron energy spectrum is uncertain due 
to the combination of factors that influence the flux density in the 
region in which spectral turnover occurs.  Taking $\alpha=0$, the circular 
polarization may be explained in 
terms of synchrotron emission from a magnetic field 
$B= 0.19 \, \xi^2 \, |\sec \theta \tan \theta | $~G, while for 
$\alpha=2$, the implied magnetic field is 
$B=0.015 \, \xi^2 \, |\sec \theta \tan \theta |$~G.  For 
$\alpha = 0$ this is equivalent to generation of circular 
polarization from electrons with an effective Lorentz factor 
$\gamma=|\cot \theta| f(\alpha)/ 3 m_c = 54.7 | \cot \theta |  \, \xi^{-1}$.
Below the self-absorption turnover frequency one has 
$T_b \approx 3.3 \times 10^{11} \, \xi^{-1} |\cot \theta|$~K, 
which is near the inverse Compton limit for $\xi^{-1} |\cot \theta| \sim 1$. 
Assuming a flux density of $640 \xi^{-1}$~mJy for the 
circularly polarized component (see Table 1), this 
brightness temperature implies an angular size of 
$0.19 \, (\cot \theta)^{-1/2}$~mas 
(1.7~AU at 8.5~kpc) at 4.8~GHz.  For $\alpha=2$ one has 
$\gamma = 193 | \cot \theta | \, \xi^{-1}$, 
$T_b \approx 1.1 \times 10^{12} \, \xi^{-1} |\cot \theta|$~K, and 
an angular size of $0.10\, (\cot \theta)^{-1/2}$~mas.  
Note that both estimates of the angular size are comparable to the 
intrinsic size 
of Sgr~A* determined by Lo et al. (1998) at $\lambda$7~mm.  
It therefore appears viable to explain the magnitude of the circular 
polarization in terms of that intrinsic to synchrotron emission.

The presence of a relativistic pair plasma has been 
suggested as the cause of circular polarization of a compact component of 
3C~279 (Wardle et al. 1998).  It is therefore 
relevant to consider the contribution 
of such a plasma to the observed properties of the circular polarization 
in Sgr~A*.

In a plasma dominated by relativistic pairs the natural modes 
of the plasma are linearly polarized.  Propagation 
through such a medium causes Stokes $U$ to cycle into $V$.  Assuming 
$V_{\rm intrinsic}=0$, and denoting the degree of linear polarization 
as $m_l$, propagation through a homogeneous medium gives rise 
to circular polarization as follows: 
\begin{eqnarray}
    m_c = m_l \sin \psi \sin \lambda^3 {\rm RRM},
\end{eqnarray}
where $\psi$ is the sky-projected angular change in magnetic field direction  
between the source region and that containing the relativistic plasma. 
The relativistic rotation measure (Kennett \& Melrose 
1998\markcite{kennett98}), 
\begin{eqnarray}
    {\rm RRM}=3 \times 10^4 L_{\rm pc} \langle n_r 
    \gamma_{\rm min} B^2 \sin^2 \theta \rangle \, {\rm rad/m}^3,
\end{eqnarray}
depends 
upon the pair density $n_r$, the path length $L_{\rm pc}$, measured in 
parsecs, and the minimum Lorentz factor of the pairs 
$\gamma_{\rm min}$.  
The linear polarization is $\sqrt{Q^2+U^2}$, and $U$ is this times 
$\sin \psi$.  Note that the linearly polarized component of 
synchrotron radiation is proportional to Stokes $Q$ only, whereas 
the relativistic plasma converts between Stokes $U$ and $V$.  The 
observed degree of circular polarization then requires ${\rm RRM} 
\approx 14 \xi/(m_l \sin \psi)$~rad/m$^3$.  \markcite{bower99a}Bower 
et al. (1999a) report
an observational limit $m_l < 0.001$.  If this reflects the degree of 
linearly polarized emission incident upon the pair-dominated region, 
one requires ${\rm RRM} > 1.4 \times 10^4 \, \xi$~rad/m$^3$ in order to 
explain the circular polarization.  
It is possible, however, that depolarization of the (presumed) linear 
polarization occurs after the partial conversion to circular polarization, in 
which case $m_l$ is higher and the corresponding limit on RRM is lower.

If linearly polarized radiation is incident upon a region containing 
an admixture of relativistic plasma and cold plasma, the ellipticity 
of the natural modes is then 
determined by the ratio $\lambda^3 {\rm RRM_m}/\lambda^2 {\rm RM_m}$, 
where RM is the rotation measure and the subscript $m$ denotes values in the 
region containing the mixture.   
The highest degree of circular polarization that can result in a 
homogeneous medium is then
\begin{eqnarray}
    m_c = m_l \sin \psi \frac{\lambda {\rm RRM}_m}{{\rm RM}_m}.
\end{eqnarray}
This is only achieved provided $\lambda^2 {\rm RM}_m \gtrsim 1$.  
In this case, the requirement on ${\rm RRM}_m$ is identical to that for a 
pair-dominated 
plasma.  However, if $\lambda^2 {\rm RM}_m \gg 1$ circular depolarization 
occurs because of rapid changes in sign with frequency.

Measurements of the circular polarization at other frequencies are required to determine 
the viability of circular repolarization models.

Finally we consider the effect of scintillation-induced circular 
polarization (Macquart \& Melrose 1999\markcite{macquart99}).  To 
exhibit this effect the source must be sufficiently small 
to undergo scintillation, and rotation measure fluctuations must be 
present in the scattering medium.   The former is likely since Sgr~A* is 
believed to exhibit variability in the total intensity 
due to interstellar scintillation (ISS) (e.g. Zhao et al. 
1993\markcite{zhao93}).  The RM fluctuations may arise from the region 
near the accretion disk (Melia 1994\markcite{Melia94} and Bower
et al. 1999a\markcite{bower99a}), or from further out in the Galactic Center 
region (e.g. Nicholls \& Gray 
1992\markcite{nicholls92}, Yusef-Zadeh, et al. 1997\markcite{yusef97}).   
The mean scintillation-induced circular polarization tends to zero only over a 
time interval large compared to the timescale of 
variability of the circular polarization.  The rotation
measure gradient required to produce the circular polarization depends
upon the variability timescale, which is related to the intrinsic
size of the scintillating source.  The timescale can influence the
expected spectral dependence of the circular polarization.  Further
observations on the variability of the circular polarization are required
to test the viability of this model and constrain the value of any
possible rotation measure gradient.

\section{Conclusion}
We confirm the detection by Bower et al. (1999b)\markcite{bower99b}
of circular polarization
from the Galactic Center source, SgrA*. We note that our detection
is from a different telescope and uses completely separate
calibration and reduction strategy.
Although clearly present, it is difficult to identify the origin of 
the circular polarization in Sgr~A*.   Measurements of the circular 
polarization over at least a decade in 
frequency are needed to test the viability of these models, 
particularly those due to synchrotron emission and circular 
repolarization. 
Measurements of any possible variability would best constrain the 
role of scintillation in producing the circular polarization.

\acknowledgments
The observations used here were
made by N.E.B. Killeen and J.-H. Zhao.  We thank R.D. Ekers, D.~Melrose and
L.~Ball for interest and encouragement with
this work, and G.C.~Bower for his comments on the manuscript.

\begin{table}
\begin{tabular}{rccccc}
\tableline
\tableline
\multicolumn{1}{c}{Date} &$I$  &$V$  &$m_{\rm V}$&RMS  &$\sigma_{\rm V}$\\
		         &(mJy)&(mJy)&(\%)       &(mJy)&(mJy)\\
\tableline
27 March 1996&616&-2.7&-0.42&0.24&0.22\\
31 March 1997&721&-2.6&-0.35&0.37&0.28\\
 1 March 1999&581&-2.0&-0.34&0.10&0.10\\
\tableline
\end{tabular}
\caption{Detections of circular polarization for Sgr~A* at 4.8 GHz \label{Table1}}
\end{table}
\end{document}